\documentclass[12pt,a4paper]{article}
\usepackage{graphicx}
\usepackage{eurosym}

\newcommand{\vv}{\vspace*{1.5ex}}

                            \newcommand{\no}{\noindent}
 \newcommand{\bc}{\begin{center}}
 \newcommand{\ec}{\end{center}}
                   \newcommand{\bfr}{\begin{flushright}}
                   \newcommand{\efr}{\end{flushright}}
   
     \newcommand{\be}{\begin{enumerate}}
     \newcommand{\ee}{\end{enumerate}}
        \newcommand{\bi}{\begin{itemize}}
        \newcommand{\ei}{\end{itemize}}
            \newcommand{\bd}{\begin{description}}
            \newcommand{\ed}{\end{description}}
                \newcommand{\beq}{\begin{equation}}
                \newcommand{\eeq}{\end{equation}}
                  \newcommand{\bea}{\begin{eqnarray}}
                  \newcommand{\eea}{\end{eqnarray}}

      \newcommand{\bfi}{\begin{figure}}
      \newcommand{\efi}{\end{figure}}
\newcommand{\bay}{\begin{array}{l}}
\newcommand{\eay}{\end{array}}
            
\makeatletter
\renewcommand\@makefntext[1]{\leftskip=2em\hskip-2em\@makefnmark#1}
\makeatother

\begin{document}   \vskip 1.5in
	\begin{center}
		{\Large {
				Universal Basic Income: The Last Bullet in the Darkness
		}}       \\[5mm]  {\large {
				Mohammad Rasoolinejad\footnote{
					Postdoctoral Research Fellow, Kellogg School of Management \\ email: mohammad.rasoolinejad@kellogg.northwestern.edu}
		}}\\
	\end{center}  \vskip 2mm   \baselineskip 13.8 pt
	
	\noindent {\bf Abstract:}\, {Universal Basic Income (UBI) has  been a controversial subject. Arguments exist on both sides in favor of and against it. Like any other financial tool, UBI can be useful if used with discretion. This paper seeks to clarify how UBI affects the economy, including how it can be beneficial. The key point is to regulate the rate of UBI based on the inflation rate. This should be done by an independent institution from the executive branch of the government. If implemented correctly, UBI can add a powerful tool to the Federal Reserve toolkit. UBI can be used to reintroduce inflation to the countries which suffer long-lasting deflationary environment. UBI has the potential to decrease the wealth disparity, decrease the national debt, increase productivity, and increase comparative advantage of the economy. UBI also can substitute the current welfare systems because of its transparency and efficiency. This article focuses more on the United States, but similar ideas can be implemented in other developed nations.
		
	}
	


\subsection*{Introduction}
Basic income is periodic cash payments to individuals with no work requirement. The basic income can be universal in means that it involves any individual in the targeted group without any test or condition and often called Universal Basic Income (UBI). If the income is uniform, every member receives the same amount of cash payment. Universality and uniformity are used interchangeably here, and both refer to the basic income which covers cash payments for every citizen. It can, however, be implemented nationally or locally. UBI may or may not satisfy the individual's basic needs. In this paper, we advocate for UBI with transient rate in which not only the basic needs might not be covered, but also the rate can fall to zero for an extended period of time. UBI differs from negative income tax, as it is not conditioned to one's income. UBI also is not the same as guaranteed income, in which the state will pay the difference if a person's annual income falls beyond a fixed threshold, in the way that UBI guarantees certain cash payments regardless of one's income or wealth.

\subsubsection*{Implementation and History}
In the early $16^{th}$ century, Thomas More depicted a Utopia in which every person receives a guaranteed income \cite{covert2018money}. Similar ideas appear in the work of Thomas Paine, Joseph Charlier and John Stuart Mill \cite{van2013universal}. UBI has generated much debate in the United States, Canada, and Europe. Several experiments have been performed during 1960s and 1970s as a negative income tax in the United States and Canada \cite{widerquist2005failure}. President Richard Nixon proposed a negative income tax bill to Congress which did not pass but approved only a minimum income for the elderly and the disabled. The Mincome experiment was held in the Manitoba province of Canada in the 1970s.  Families received an income at a reduced rate with respect to earned money by wage \cite{hum1993economic}. In the 1980s, basic income was more or less forgotten in the United States but gained some traction in Europe. 

Pilot programs were tried in countries such as Namibia, Kenya, Brazil, Finland, Canada, and India similar to UBI. In 2016 Switzerland held a national referendum on basic income \cite{straubhaar2017economics}. The idea was rejected by a vote of 76.9\% to 23.1\%. Finland launched a two-year pilot program in 2017 in which 2,000 unemployed individuals received \euro{560} per month. Preliminary results show not much change in unemployment, but of the recipients showed a reduction in stress and anxiety during the period in which they received basic income \cite{kangas2019basic}. 

Alaska's Permanent Fund can be categorized as a type of basic income, in that the state distributes the revenue generated from the oil and gas industry \cite{goldsmith2010alaska}. In 2017, Ontario launched a three-year basic income program in which 4,000 low-income participants received an annual stipend (individuals received \$13,000, and couples received \$19,000) \cite{de2018rise}. The government charged 50 cents for every dollar earned through work. The recipients were also required to opt out of some social programs. The program was canceled after one year due to extraordinary costs to taxpayers. Another project in Stockton, California started in February 2019 for duration of 18 month which gave 100 low-income families \$500 per month \cite{hamilton2019universal}. The initial results show people spent most of the incomes on basic needs such as food, clothing and utility bills. 



\subsection*{UBI and Inflation}
UBI can be funded through money printing, deficit spending or by revenue from taxes. In the first case, the base money supply will increase. The risk with this case is raising inflation. As long as inflation is under control, money printing can help in many other issues (such as alleviating national debt or increasing comparative advantage of country in the trade) in a manner that is fair to everyone. The following sections will elaborate more on the benefits that money printing can provide.

The government will issue the notes to fund the UBI program in deficit spending. This case resembles transferring money from the wealthy (local or foreign) to the middle class and poor. However, the lower and middle class become endebted to the wealthy and their debt must be paid at a later time. Deficit spending may also cause inflation, as lower-income households tend to spend a higher percentage of household income. This method is not preferable, as raising money through debt issuing is not sustainable in the long run. 

Funding UBI with tax income is also transferring money from the wealthy to the middle class and poor, except this time the money will not be paid back. UBI can be funded through land/location value tax (LVT), value-added tax (VAT) or income tax. Some have also suggested the use of Robot-tax, which penalizes industries for the use of automation instead of human labor \cite{desai2019some}. To punish industries for technological advancements, however, is irrational. This will reduce the comparative advantage of local industries. The disadvantage with income tax is that raising taxes on the wealthy usually hinders economic growth. Between the suggested taxation methods, the VAT seems the best option, as the program is funded by the individuals who spend the most, and it encourages people to save. Here and through the rest of the paper, we assume UBI is funded by money printing. The author believes other methods will decrease or eradicate the purpose of implementing UBI in the first place.

In theory, printing money and increasing the money supply causes inflation. Many leading world economies such as Japan and the Europe Union currently suffer from a deflationary environment. Japan has struggled to introduce inflation into the economy for a couple of years. Even massive monetary easing and asset purchasing do not seem to work efficiently. Such an environment also more or less exists through the Europe Union. Many reasons cause these advanced economies to face a deflationary environment. From them one can mention demographics and reduction of births, wealth disparity, lack of spending, lagging in wage growth, increase in national and domestic debt, low manufacturing costs due to automation, the rise of the Internet and e-commerce, increase in competition, electronic banking, close loop inflation control by the Federal Reserve in the US and other central banks around the world, long-lasting low inflation environment and low inflation expectations, the 2008 financial crisis, etc. Many of these forces do not seem to go away in the foreseeable future.

Money in the hands of the wealthy tends to increase the price of financial assets, whereas money in the hands of the middle class will drive spending on common goods and will cause inflation. The traditional monetary easing and asset purchasing are not the most effective ways to produce inflation. The radical measures implemented by central banks around the world to increase economic productivity and inflation seems not to work anymore. Inflation in security prices was already caused by decreasing the interest rate to their lower bound. The inflation in the bond market keeps the interest rates low, which makes the window of potential further cut the interest rates narrow. These methods increase the monetary supply; at the same time, however, wealth disparity also increases. Corporations and the wealthy benefit the most from such methods. UBI, on the other hand, increases the money supply and at the same time decreases wealth disparity. The spending driven by the middle class will cause rise in prices.

UBI has failed in many cases solely due to an uncontrolled money supply to the middle class, which produced hyperinflation. The catch here is that UBI should not be controlled by the executive branch of government. In the case of democracy, it is quite logical to assume each candidate will promise higher UBI rates to gain more votes. This will produce a vicious cycle ending with hyperinflation. UBI should be under the control of an independent entity, such as the Federal Reserve in the United States. That entity will control the money supply and amount of UBI based on projections of inflation. UBI will be an extra tool of the Federal Reserve which can directly target inflation and wealth disparity.  

As an example assume the target inflation to be two percent and country's inflation rate sustainably runs below the target. Eq. \ref{e2} gives an estimation for the UBI rate to be paid to achieve the target inflation. After running the program for certain amount of time (for example three months), the inflation rate shall be reassessed. If running below or above the target range, the rate can be readjusted accordingly. UBI rate eventually goes to zero if the inflation rate sustainably runs above the target range.

UBI works best in a deflationary environment, not in an inflationary one. It is best suited for developed countries with deflationary or low inflation environment. In such countries, not only inflation to be avoided, but the central banks are pushing to raise inflation to a healthy level. For developing countries often with higher inflation rates, UBI can cause disastrous effects.

Some have suggested that UBI can be implemented for the currency inflation \cite{standing2017basic,standing2016corruption}. The UBI is not good to compensate for inflation, as it is itself an inducer of inflation. In this case, more inflation would cause increase in UBI rates, which makes the inflation problem worse. Also, the amount one loses due to inflation is proportional to his/her cash holding, so people with more cash should receive more basic income, which does not seem rational.

\subsection*{UBI and Wealth Disparity}
Current society in developed countries suffers from inequality and a huge wealth disparity. Such an environment leads to the dissatisfaction of the poor and middle class, which can lead to crisis and chaos. Examples of this also happened prior to World War II. The wealth disparity can be decreased by social programs which are usually funded by taxes. Taxation, and especially progressive taxation, seems unfair to the wealthy and also acts as decelerator for the economy. Corporate tax tends to decrease the profit margin of corporations and decreases their comparative advantage on the global scale.

As mentioned in the last section, the traditional monetary easing and asset purchasing barely introduced any inflation as most of the money got absorbed by financial markets and increased the wealth of the wealthy. UBI, on the other hand, does not treat the wealthy differently and is fair to everybody, as each person receives the same amount of money. The same payment increases more the wealth of the poor and middle class in relative terms. So, in a deflationary or low-inflation environment (which might be caused by wealth disparity itself), the equal UBI payment to everybody tends to decrease the wealth disparity and Gini coefficient. The new Gini coefficient is calculated as:

\beq \label{e1}
G^{new} = \frac {M_3} {M_3 + U} \ G^{old}
\eeq
which $G^{old}$ is the Gini coefficient before UBI payment, the $M_3$ is M3 monetary supply that is an indicator of the amount of wealth in society and $U$ is total (not rate) cash payments as basic income to all individuals (this formula assumes universality and uniformity). The inflation induced by UBI should be proportional to 

\beq \label{e2}
\delta \pi \approx \frac {UV} {M_1}
\eeq
in which $V$ is velocity of money and $M_1$ is M1 monetary supply.

\subsection*{UBI and Work Desirability}
There are many concerns that UBI will lead to shrinkage in the workforce. One should note, however, that the rate of UBI should be at its maximum in a deflationary environment (when unemployment is high). As the economy progresses and the job market becomes tight, wages and inflation will rise. This should cause the rate of UBI to decrease or even go to zero. So, more people will depend on the job for income. UBI will act as a regulator; the workforce will be taken out of the market when they are not needed and will be reintroduced when they are in need.

If the economy is good but inflation is missing (such as in 2019), providing UBI may curb off part of the job market. This would cause an increase in wages and also prevents the economy from overheating. The wealthy would have to share much of their profits with the lower part of society. The increase in wages will eventually cause inflation, and return of the workforce. At the same time, rising inflation reduces basic income rate, and the market should then return to equilibrium.

Some argue that UBI payments will lower incentive to work and singlehandedly cause a decrease in the workforce \cite{tanner2015pros,birnbaum2008just}. The argument seems false as low employment will cause wages to rise, and eventually, the marginal benefit for the worker forces him to return to the job market. It should be obvious that most people in the current environment do not work just to cover their basic needs. The only way that will cause a lack of incentive to work is to set a minimum guaranteed income. This means that regardless of whether a person works, he/she will always receive a minimum amount of money.

In the UBI system advocated here, everybody in society receives a certain amount of money. If a person decides to work, he will be paid separately for his work.  Therefore, there is no way that the person who works will have the same income as the person who does not. The marginal benefit (extra income) will produce marginal work desirability. An experiment which was done in Dauphin, Manitoba in 1970 also indicates that the argument is false and basic income does not cause mass unemployment \cite{forget2011town}. One study on the Alaska Permanent Fund Dividend revealed that while a small decrease in labor force observed, the part-time jobs increased 17 percent \cite{jones2018labor}. Another study conducted in the Namibian village of Omitara found that the economic activity increased by the introduction of basic income especially in small businesses \cite{jones2018labor}. While some labor markets might lose workforce, higher spending supports overall employment. It has to be noted that it is possible to deteriorate the economy by producing hyperinflation, and this can hurt the job market. The rate of UBI should be set to prevent a high inflation rate.

It is improbable in the foreseeable future for humankind to become completely self-sufficient. What machines are incapable of, is to substitute for human emotions. Our needs may shift from basic survival needs to others. The skill sets we now consider most important might be substantially different in the next era. The need will come for certain skills, and people will pay for them. This transition in demand for skills has happened multiple times throughout human history, and this time will not be the last. Regulated UBI should lead to an increase in productivity and creativity. If people are sure to have support, especially in the case of an economic downturn, they will be freer to pursue jobs about which they are passionate. This would lead to a substantial increase in productivity, making room for entrepreneurship and innovation. 

\subsection*{UBI and Recession}
A recession is defined as the decline of corporate profits for two consecutive quarters \cite{abberger2008define}. The rate of defaults increases as debtors' profits can not sustain the current debt. Credit dries up as the creditors become more cautious in lending. A recession is detrimental to the middle class, as corporations cut jobs to reduce the cost. This also further decreases the spending power of the middle class, creating a vicious cycle. Banks are private entities and tend not to give out money to individuals with low credit in a low-interest environment. This will cause most of the available credit due to lowering interest rates to go to corporations and wealthy people. Money in the hands of the wealthy tends not to increase the velocity of money very much, with respect to when money ends up in the hand of the middle class. An increase in money velocity can help to revive the economy and bring back inflation.

In past recessions, the Federal Reserve usually interfered by decreasing interest rates, asset purchasing, and monetary easing. This helped debtors to refinance the old debt with new cheap credit. In the next recession, there will be not much room for cuts in the interest rate, as the rates are already close to the lower bound. The cheap access to credit during the last decade caused a boom in the bond market and most of the credit consumed for buying the low return assets. In this situation, UBI can come to the rescue. The benefits of UBI in a recessionary situation are: (1) It increases the money supply regardless of the lower bound in interest rates. (2) It is very effective for increasing the middle class's spending power, which drives corporate profits. (3) It pushes up the long-term interest rates as inflation rises which will make more room for cuts in interest in the next recessions. (4) It works best in a deflationary environment.

\subsection*{UBI and Unemployment}
Recession-induced unemployment and how the UBI helps to recover the corporate profits which eventually leads to a fall in unemployment have already been discussed. There, unemployment is not systematic and the workforce returns to the market as the economy peaks. On the other hand, the rise of Automation and Artificial Intelligence (AI) can cause massive systematic unemployment. For example, the development of autonomous vehicles can lead to huge unemployment in the transportation sector. Congress reported in 2010 that there is a high probability that most low-skilled jobs will be replaced by computers. The rise of the Internet and e-commerce and a low number of human capital-intense businesses generate further concerns. There is a high probability that a huge wave of unemployment will come, especially in the less-educated part of society.

Unemployment will lead to a loss of earning power and a decrease in spending. Wealth disparity will rise to huge levels. The wealthy will become much wealthier as production no longer needs the cost of human labor. The economy might go into a deflationary mode for the following reasons: (1) Lack of spending by the middle class due to unemployment and lack of income. (2) The amount of production will increase. (3) The cost of producing also falls substantially. These factors result in falling prices and induce deflation. In such an environment, UBI can maintain the purchasing power of society, produce an inflationary environment and decrease wealth disparity. UBI is efficient especially in a low production cost environment, as payments will cover more of the recipients' needs.

\subsection*{UBI, National Debt and Trade Deficit}

Currently, the United States and most advanced countries suffer from a huge amount of national debt. The low-interest-rate environment has led to the accumulation of huge debt, which in some cases grew multiple times the size of a country's Gross Domestic Product (GDP). It seems improbable that such countries can pay their debt in real term. UBI can produce inflationary forces, devalue the currency and ease the debt burden. For many years the strong dollar and high labor cost reduced the comparative advantage of the US economy relative to the other global industries. The devaluation of currency makes the country more competitive in the global term and causes the trade deficit to decrease. 

\subsection*{Welfare and Poverty Reduction}
The current welfare system is inefficient, costly, and unfair. The United States spends more than two trillion dollars on Medicare and social security annually. The high cost of entitlements reduces capital investment and economic growth \cite{greenspan2014map}. This system of welfare is not transparent and wastes a lot of capital on administrative and bureaucratic costs. The ideal welfare system must be low cost, easy to handle, and fair to every citizen.

Fairness is an important factor that is often neglected. The welfare system should not systematically discriminate against any susceptible group from another. For example, forgiving student debt is not fair, as the recipients receive a different amount of treatment depending on their debt. Instead of rewarding the one who selected not to amass debt or finished paying their debt, we punish them by forgiving the ones who acted irresponsibly. One reason why most social programs fail is that they neglect the fairness at the core.

UBI may be considered a welfare system that can eventually replace current ones \cite{ravallion2015economics}. Separate welfare programs can be replaced by a general UBI program. Universality and uniformity guarantee fairness in that each person receives the same amount of money. Besides that, uniform income is much less bureaucratic, easier to implement, and has less administrative cost to run \cite{standing2008cash,bregman2017utopian,standing2017basic}. It substantially reduces the cost of assessment and also prevents the exclusion and inclusion errors \cite{hanna2018universal}. Universal income is more transparent and is less susceptible to corruption. 

Some have suggested that social dividend can be used as a new welfare system \cite{marangos2004social,russell1919proposed}. In this regard, social dividend \cite{marangos2004social} falls short of UBI, as it is more prone to corruption and lacks efficiency. The social dividend will be highest in an economic boom and lowest during a recession (when they are needed most). The social dividend can not be used to regulate inflation. These examples of market socialism have been implemented before and have mostly failed.  

UBI should not be used as a tool to gather more votes in democratic countries. Income from welfare should be run under an independent entity working closely with both the executive branch of government and the Federal Reserve. The rates can be assigned to each susceptible group; these criteria, however, should be easy to figure out, such as age, to prevent inefficiency and bureaucracy. Each susceptible group can then use their share of basic income to pay the premium for insurance to reduce the risk in the specific matter to which he/she is more susceptible. This method guarantees fairness and ensures that the funds are not targeted to a specific group.

UBI assigned to children under the legal age can either go to their parents or be collected in a trust fund to become available when they become of age. If the payment goes to parents, that could help to cover some cost of child-raising. It would not likely help much to incentivize the childbearing as the rate probably will not cover all the costs; but anyway, the minimal effects are welcome in the current environment as most countries with low inflation also suffer from a low rate of childbirth.

Advocates of basic income often argue that it could potentially reduce or even eradicate poverty \cite{bregman2017utopian}. The poor benefit the most from UBI, as their purchasing power increases the most relative to their income. More spending by the poor and middle class results in more corporate profit. The increase in corporate profits benefits the wealthy and the country as a whole. Some argue that people will spend their basic income on drugs and alcohol. This argument is false in the sense that there will be good and bad use for any mean. The studies performed on the effect of basic income also do not confirm the argument \cite{evans2017cash}. Even if these effects appear, they will be local as people tend to improve their situations in time, especially through generations.

If UBI covers basic survival needs, people will have the freedom to choose their path and pursue their passions. This will increase the power of the middle class and help to increase social mobility. The way UBI improves the lower part of society is not only in terms of wealth disparity but it also increases the power of employees in the labor market. UBI increases social insurance and reduces insecurity which means more bargaining power for the middle and poorer class. It also reduces the stress and anxiety of society. One might note, however, that the minimum needs will not control the UBI rate but the inflation, if we want to continue the program sustainably. It should not generate the notion that UBI will be paid in any condition and will cover basic needs. It will work like the interest rates as the Federal Reserve changes it based on its discretion.

\subsection*{Summary and Conclusions}
We discussed how the struggles in leading economies (such as wealth disparity, lack of inflation, low births, etc.) are tied together and how UBI can offer a solution for them, if it is implemented in a controlled manner. That necessitates regulation from an independent institution that sets the UBI rate based on the target inflation. UBI works best in a deflationary environment and should not be used as a fixed amount to compensate for the cost of living. It can raise inflation, decrease inequality, and alleviate national debt. UBI also has the potential to offer a more transparent and efficient welfare system. 

\vv \no {\bf Acknowledgment:}\ {Thanks to Mr. Samuel B. Cotten for proofreading and editing the manuscript.}
	
	%
	%

	\addcontentsline{toc}{section}{References}
	\bibliographystyle{unsrt}
	\bibliography{UBI}

	
\end{document}